\documentclass[aps, prl, twocolumn, superscriptaddress, floatfix, notitlepage]{revtex4-2}
\usepackage{amsbsy}
\usepackage{amsfonts}
\usepackage{amsmath}
\usepackage{amstext}
\usepackage{amsthm}
\usepackage{amssymb}

\usepackage{graphicx}

\usepackage{epsfig}
\usepackage{epstopdf}
\usepackage{bm}
\usepackage{color}
\usepackage{xcolor}
\definecolor{prlblue}{rgb}{0.18,0.19,0.57}
\usepackage{multirow}
\usepackage[colorlinks]{hyperref}
\usepackage{cleveref}
\hypersetup{citecolor=prlblue, linkcolor=prlblue, filecolor=prlblue, urlcolor=prlblue}
\newcommand{\figref}[1]{Fig.\,\ref{#1}}

\newcommand{\eqnref}[1]{Eq.\,\eqref{#1}}

\begin{document}
\graphicspath{{/}}

\title{Strong Pairing Originated from an Emergent $\mathbb{Z}_2$ Berry Phase in La$_3$Ni$_2$O$_7$}

\author{Jia-Xin Zhang}
\thanks{These two authors contributed equally to this work}
\email{zhangjx.phy@gmail.com}
\affiliation{Institute for Advanced Study, Tsinghua University, Beijing 100084, China}

\author{Hao-Kai Zhang}
\thanks{These two authors contributed equally to this work}
\email{zhk20@mails.tsinghua.edu.cn}
\affiliation{Institute for Advanced Study, Tsinghua University, Beijing 100084, China}

\author{Yi-Zhuang You}
\affiliation{Department of Physics, University of California, San Diego, CA 92093, USA}

\author{Zheng-Yu Weng}
\affiliation{Institute for Advanced Study, Tsinghua University, Beijing 100084, China}

\date{\today}

\begin{abstract}
The recent discovery of high-temperature superconductivity in La$_3$Ni$_2$O$_7$ offers a fresh platform for exploring unconventional pairing mechanisms. Starting with the basic argument that the electrons in $d_{z^2}$ orbitals nearly form local moments, we examine the effect of the Hubbard interaction $U$ on the binding strength of Cooper pairs based on a single-orbital bilayer model with intralayer hopping $t_{\|}$ and interlayer super-exchange $J_{\perp}$. By extensive density matrix renormalization group calculations, we observe a remarkable enhancement in binding energy as much as $10$-$20$ times larger with $U/t_\|$ increasing from $0$ to $12$ at $J_{\perp}/t_\|\sim 1$. We demonstrate that such a substantial enhancement stems from a kinetic-energy-driven mechanism. Specifically, a $\mathbb{Z}_2$ Berry phase will emerge at large $U$ due to the Hilbert space restriction (Mottness), which strongly suppresses the mobility of single particle propagation as compared to $U=0$. However, the kinetic energy of the electrons (holes) can be greatly restored by forming an interlayer spin-singlet pairing, which naturally results in a superconducting state even for relatively small $J_\perp$. An effective hard-core bosonic model is further proposed to estimate the superconducting transition temperature at the mean-field level.
\end{abstract}

\maketitle


{\it Introduction.---}Ever since the revelation of high-temperature superconductivity (SC) in cuprates~\cite{Tranquada.Fradkin.2015doan, Zaanen.Keimer.2015, Wen.Lee.2006z4} -- commonly recognized as a doped Mott insulator -- the quest to understand the relationship between the pairing mechanism of unconventional SC and strong electron correlations has persisted as an enduring challenge~\cite{Wen.Lee.2006z4, Scalapino.Scalapino.2012}. The recent experimental breakthrough~\cite{Wang.Sun.2023, Yuan.Zhang.2023, Cheng.Hou.2023, Wen.Liu.2023}, revealing high-temperature superconductivity in pressurized single crystals of La$_3$Ni$_2$O$_7$ (LNO), has aroused broad interest. With an observed maximum SC transition temperature $T_c$ reaching 80K under pressures exceeding 14GPa~\cite{Wang.Sun.2023}, LNO presents a new platform to delve into and scrutinize unconventional pairing mechanisms.

LNO features a layered structure wherein each unit cell incorporates two conductive $\mathrm{NiO}_2$ layers, paralleling the $\mathrm{CuO}_2$ layer found in cuprates. Insights drawn from density functional theory (DFT)~\cite{Wang.Sun.2023, Leonov.Shilenko.2023, Kuroki.Sakakibara.2023, Werner.Christiansson.2023, Dagotto.Zhang.2023, Dagotto.Zhang.2023b, Dagotto.Zhang.2023c, ZhangHu2024} calculation suggest that the low-energy behaviors in LNO are dominated by two $e_g$ orbitals of $\mathrm{Ni}$, namely $3d_{x^2-y^2}$ and $3d_{z^2}$ with the filling $\nu \approx 1/4$ and $\nu \approx 1/2$, respectively, which has been further verified by Hall coefficient transport measurements~\cite{doi:10.1143/JPSJ.64.1644, doi:10.1143/JPSJ.65.3978, wang2024normal}, as well as the volume counting of the Fermi surface observed in Angle-resolved photoemission spectroscopy (ARPES)~\cite{2024NatCo..15.4373Y, ZhangWen2024}. When under pressure, the interlayer Ni-O-Ni bonding angle changes from $168^{\circ}$ to a straightened $180^{\circ}$, which significantly enhances the interlayer coupling. Furthermore, the pronounced Coulomb repulsion within the Ni-3d orbitals merits attention, aligning with the latest experimental data that posits LNO as nearing a Mott phase and exhibiting non-Fermi-liquid traits above $T_c$, marked by a linearly temperature-dependent resistivity that extends up to 300K~\cite{Yuan.Zhang.2023, Wen.Liu.2023,Wang.Sun.2023}.

Two prevalent theoretical starting points exist for describing the SC pairing mechanism in LNO. One relies on a weak coupling approach~\cite{Kuroki.Sakakibara.2023,Wang.Yang.2023, Hu.Gu.2023, Eremin.Lechermann.2023, Yang.Liu.2023, You.Lu.2023}, attributing the SC pairing to the instability of Fermi pockets with spin fluctuations as the pairing glue. Alternatively, the strong-coupling perspective~\cite{Kuroki.Sakakibara.2023,Wang.Yang.2023, Hu.Gu.2023,Su.Qu.2023, Wu.Lu.2023kwo, Oh2023, Zhang.Jiang.2023, Zhang.Yang.2023, Wang.Wu.2023, Zhang.Shen.2023, Si.Liao.2023, You.Lu.2023} emphasizes the role of interlayer exchange between the $d_{z^2}$ orbitals within a unit cell. Regardless of the approach taken, many works point out that it is the spin exchange interaction, which is the residual effect of Coulomb repulsion (on-site Hubbard repulsion), rather than the Coulomb repulsion itself, to be considered as the key driver in the pairing mechanism. This focus may arise from a clear difference between LNO and cuprates: the doped hole density in the $d_{x^2-y^2}$ orbital for LNO ($\delta=2(1-\nu)\approx 0.5$), is significantly larger than that in cuprates. In cuprate systems~\cite{Tranquada.Fradkin.2015doan, Zaanen.Keimer.2015}, such high doping levels correspond to the ``Fermi-liquid'' phase, seen as the breakdown of the ``Mottness''. As a result, one might initially consider the Coulomb repulsion in the $d_{x^2-y^2}$ orbital of LNO to be irrelevant. If that is the case, such high $T_c$ as observed in experimental data would necessitate a relatively dominant spin-exchange interaction to facilitate the formation of robust Copper pairs.

\begin{figure}[t]
    \centering
    \includegraphics[width=\linewidth]{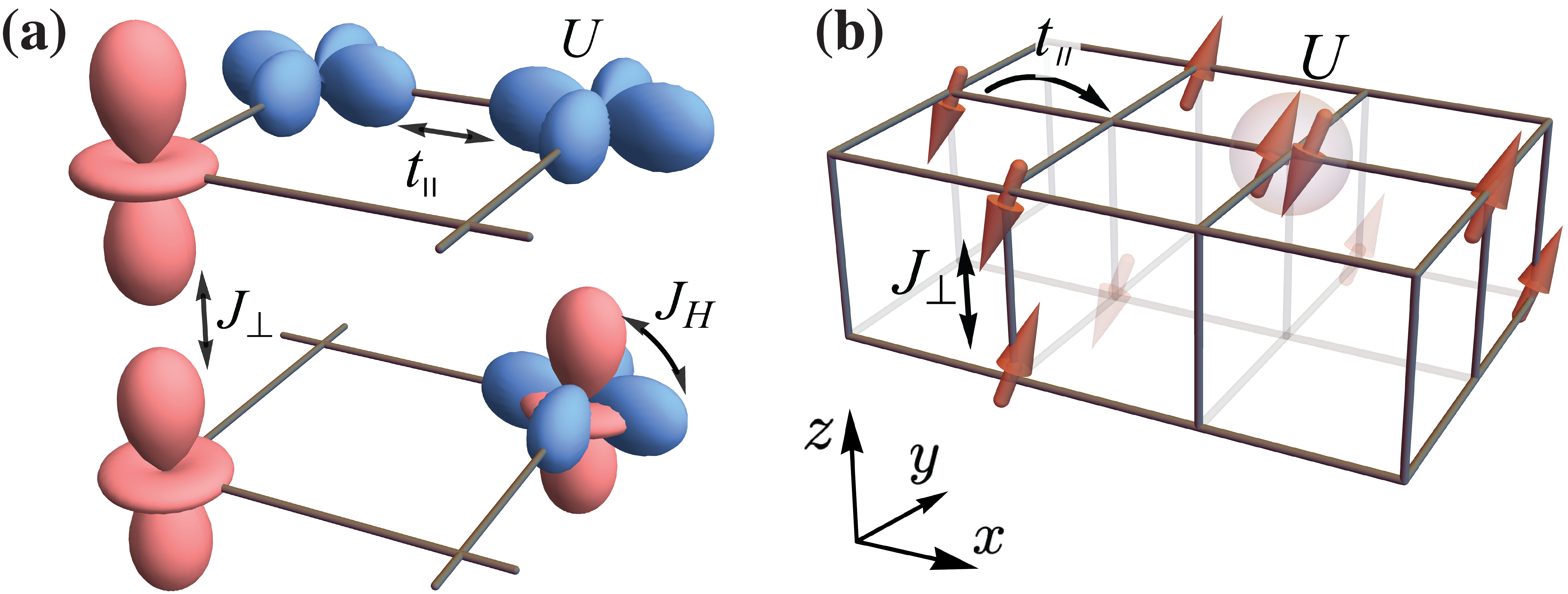}
    \caption{(a) Schematic representation of the low-lying two-orbital physics in LNO: the red $d_{z^2}$ orbital displays interlayer spin exchange coupling $J_\perp$, while the blue $d_{x^2-y^2}$ orbital features intralayer hopping term $t_{\|}$ as well as on-site Hubbard interaction $U$. The Hund's coupling $J_H$ links these two orbitals. (b) Minimal model for $d_{z^2}$ orbital, applicable when $J_H$ is strong enough to align the spins between the $d_{x^2-y^2}$ and $d_{z^2}$ orbitals.}
    \label{fig_orbital}
\end{figure}

\begin{figure}[t]
    \centering
    \includegraphics[width=0.92\linewidth]{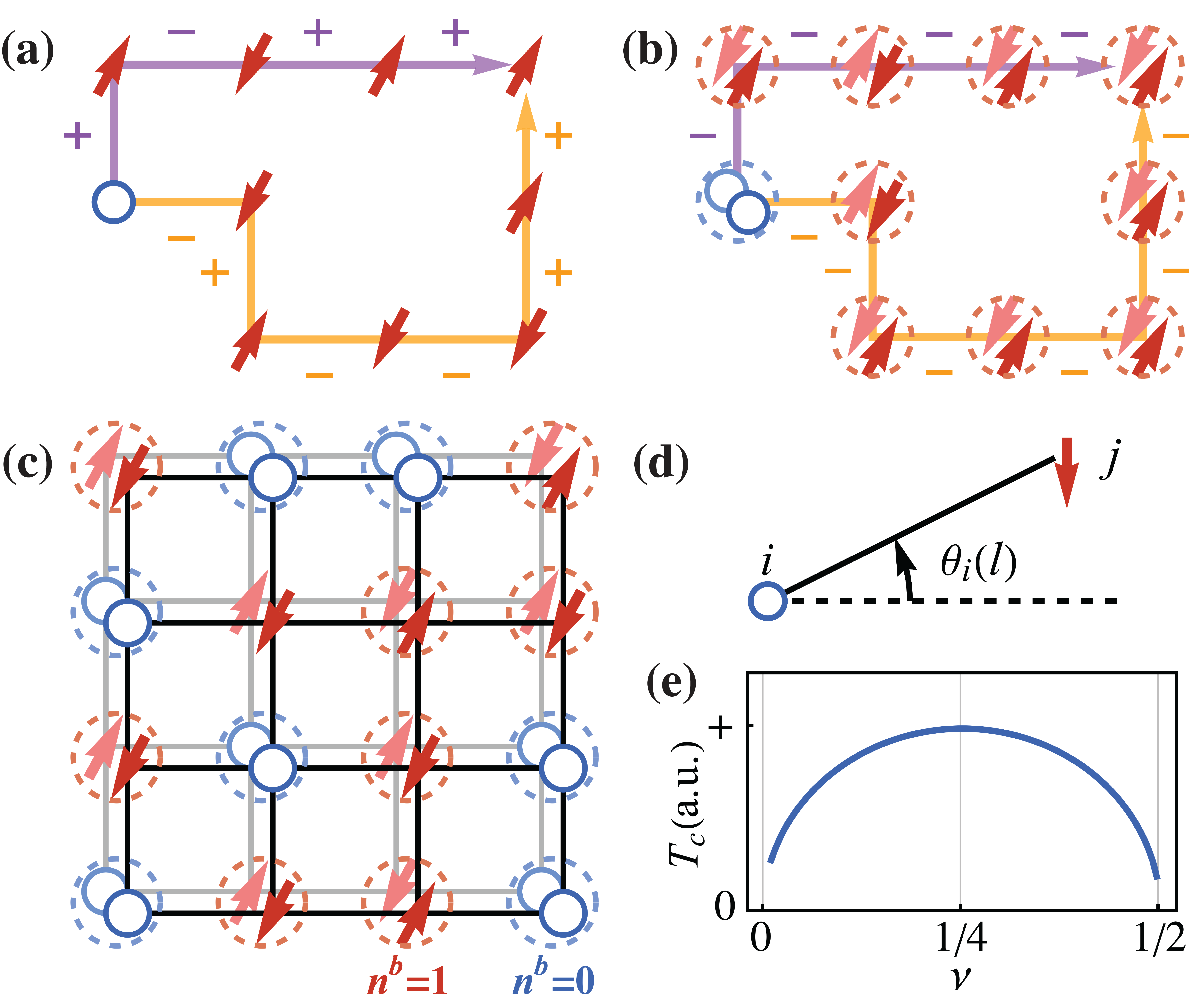}
    \caption{Schematics of the $\mathbb{Z}_2$ Berry phase experienced by (a) one hole and (b) hole pairs in the spin-singlet background. Dark (light) blue circles indicate holes in the 1st (2nd) layer, while dark (light) red arrows represent spins in the 1st (2nd) layer. (c) Schematic of the effective hard-core bosonic model in \eqnref{Heff}: Spin-singlet pairs correspond to $n^b_i=1$ and hole pairs correspond to $n^b_i=0$. (d) Graphic representation of $e^{i \hat{\Theta}}$ in \eqnref{unitary} in the two-dimensional case. (e) Superconducting $T_c$ vs filling $\nu$ obtained by the effective model \eqnref{Heff} at the mean-field level, with $J_{\perp}=0.6 t_{\|}$ and $J_{\|}=2J_{\perp}/3$.}
    \label{fig_illu}
\end{figure}

However, as we will show, the density matrix renormalization group (DMRG) results indicate that the on-site Hubbard interaction indeed substantially enhances the pairing strength, particularly when the intermediate interlayer spin exchange coupling is close to estimation from DFT calculations~\cite{Wang.Sun.2023, Leonov.Shilenko.2023, Kuroki.Sakakibara.2023, Werner.Christiansson.2023, Dagotto.Zhang.2023, ZhangHu2024}. This numerical evidence offers a compelling hint that the Coulomb interaction itself still plays a significant role in the pairing mechanism. Based on this, we will further demonstrate that the motion of a single hole is severely restricted due to the accumulation of an emergent $\mathbb{Z}_2$ Berry phase, which arises from the restricted Hilbert space due to the strong on-site Hubbard interaction. Only the channel of inter-layer hole pairs on the background of inter-layer singlet spin pairs remains unvanished weight because the $\mathbb{Z}_2$ Berry phase can be fully canceled [shown in \figref{fig_illu}(b)]. This kinetic-energy-driven pairing mechanism originates directly from the strong Coulomb repulsion, which is applicable across a wide range of hole doping concentrations and is not sensitive to other orbital specifics, such as particular inter-orbital couplings.

{\it Binding energy from Hubbard repulsion.---}We first analyze LNO and establish our minimal model. The electronic correlation in the nearly half-filled $d_{z^2}$ orbitals is remarkably strong, as supported by experiments showing that the bands mainly contributed by $d_{z^2}$ orbitals lie below the Fermi surface, are much flatter, and exhibit strong band renormalization (5-8 times) compared with the bands contributed by $d_{x^2-y^2}$ orbitals~\cite{2024NatCo..15.4373Y, LiYang2024}. Additionally, numerical calculations including DMRG~\cite{LiSu2023} and dynamical mean-field theory (DMFT)~\cite{2024HeLu} that consider multiple orbitals demonstrate that charge carriers are primarily contributed by $d_{x^2-y^2}$ orbitals, rather than $d_{z^2}$ orbitals, and that inter-orbital Hund's coupling plays an important role in the formation of superconductivity.

Therefore, it is reasonable to consider a scenario in which the majority of electrons in the nearly half-filled $d_{z^2}$ orbitals tend to localize, forming local magnetic moments, while their charge degrees of freedom become nearly frozen. This phenomenon can be attributed to the emergence of an orbital-selective Mott transition, predominantly driven by Hund's coupling among the distinct $d$-orbitals~\cite{PhysRevLett.102.126401, PhysRevLett.112.177001, 125045}. Consequently, as shown in \figref{fig_orbital}(a), the only relevant interactions are the interlayer spin exchange interaction $J_{\perp}$ in $d_{z^2}$ orbitals, the intralayer hopping $t_{\|}$ and on-site Coulomb interaction $U$ in $d_{x^2-y^2}$ orbitals (the $d_{x^2-y^2}$ orbital exhibits negligible interlayer single-particle tunneling), as well as the Hund's coupling $J_H$ between different orbitals. Here $U$ and $J_\perp$ are treated as independent parameters to investigate the effect of Mottness. Other allowed complicated interactions in generic multi-orbital systems~\cite{125045}, such as interorbital hopping, are renormalized to be very small due to strong correlation effects~\cite{202310YangWu}. The inter-orbital Coulomb interaction effectively only provides a chemical potential shift to the $d_{x^2-y^2}$ orbitals due to the nearly fixed particle number of $d_{z^2}$ orbitals.

Notably, the strong $J_H$ tends to polarize the spin in different orbitals so that in the large $J_H$ limit, the interlayer exchange interaction $J_\perp$ for $d_{z^2}$ orbital can transfer to $d_{x^2-y^2}$ orbital~\cite{Wu.Lu.2023kwo} after integrating out the $d_{z^2}$ local moments. Therefore, we propose the single-band minimal model describing the $d_{x^2-y^2}$ orbital with filling $\nu \approx 1/4$, as specified in [cf.~\figref{fig_orbital}(b)]:
\begin{eqnarray}\label{H1}
    H_{t_{\|}\text{-}U\text{-}J_{\perp}}&=&-t_{\|} \sum_{\langle ij\rangle \alpha \sigma}\left(c_{i\alpha \sigma}^{\dagger} c_{j \alpha \sigma}+\text { h.c.}\right) - \mu \sum_{i\alpha\sigma}n_{i\alpha\sigma}\notag\\
    &\;&+U \sum_{i \alpha} n_{i \alpha \uparrow} n_{i\alpha \downarrow} +J_{\perp} \sum_i \boldsymbol{S}_{i 1} \cdot \boldsymbol{S}_{i 2},
\end{eqnarray}
where $\sigma=\uparrow, \downarrow$ is the spin orientation, and $\alpha=1,2$ denotes the layer index. $\boldsymbol{S}_{i\alpha}$ and $n_{i\alpha \sigma}=c_{i \alpha\sigma}^{\dagger} c_{i \alpha \sigma}$ are the local spin operator and electron number operator, respectively. We set the intralayer hopping $t_\| = 1$ as the energy unit. The interlayer spin coupling $J_\perp$ and the Hubbard repulsion $U$ are varied independently to investigate the effect of Mottness.

\begin{figure}
    \centering
    \includegraphics[width=0.99\linewidth]{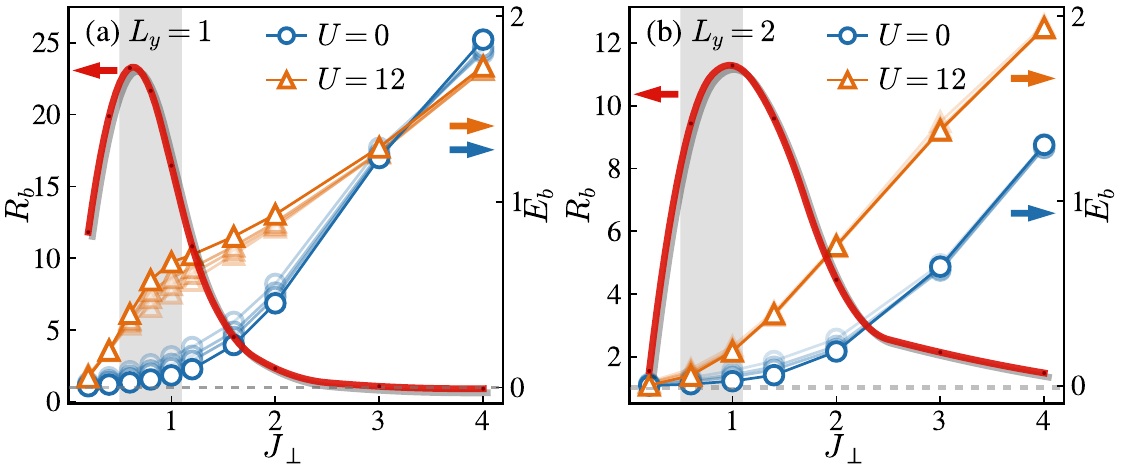}
    \caption{The binding energy $E_b$ for different on-site repulsion $U$ and the corresponding ratio $R_b$ as a function of interlayer coupling $J_\perp$ with $L_y=1,2$ for panel (a) and (b), respectively. The system length is $L_x=8,12,16,32$ for markers of decreasing transparency. The hole doping is fixed at $\delta=1/2$. The horizontal dashed lines mark $R_b=1$ and $E_b=0$. The shaded areas indicate the experimental relevant regions of $J_\perp$.}
    \label{fig:hubbard_binding_vary_jp}
\end{figure}

We perform DMRG calculations on a bilayer lattice of size $L_x\times L_y$ with $L_y$ being the shorter side. The total number of sites is $N=L_x\times L_y\times L_z$ with $L_z=2$. We keep the bond dimension up to $5000$ with a typical truncation error $\epsilon\lesssim 10^{-7}$. The binding energy of holes is defined as
\begin{equation}\label{Eq:Eb}
\begin{aligned}
    E_{b} = 2E(N_h+1)-E(N_h+2)-E(N_h),
\end{aligned}
\end{equation}
where $E(N_h)$ denotes the ground state energy with the fixed number of doped holes $N_h=N\delta$. The total spin $S^z$ is fixed at $0$ and $1/2$ for even and odd numbers of holes respectively. Note that the binding energy has an overall linear dependence of $J_\perp$ since the Hamiltonian scales with $J_\perp$. Hence, we define a dimensionless ratio
\begin{equation}
    R_b = \frac{E_b(U=12)}{E_b(U=0)},
\end{equation}
to reflect the enhancement of binding energy caused by the Hubbard repulsion. The value of $R_b$ for different $J_\perp$ is depicted by the red lines in \figref{fig:hubbard_binding_vary_jp}, which shows a significant enhancement in binding energy due to on-site repulsion $U$ in both cases of $L_y=1$ and $L_y=2$. Notably, this effect is most pronounced when the interlayer coupling $J_\perp$ is relatively small, specifically when $J_\perp\lesssim t_\parallel$, which is in line with the realistic values obtained through DFT calculations. Certainly, such a huge amplification of binding energy still necessitates the assistance of a finite interlayer coupling, albeit not excessively so. This is because at doping $\delta=1/2$, neither the Hubbard chain nor the two-leg Hubbard ladder is a quasi-one-dimensional superconducting state (as the Luther-Emery liquid), which is consistent with the vanishing binding energy at $J_\perp\rightarrow0$ shown in \figref{fig:hubbard_binding_vary_jp}. One is referred to Supplemental Material~\footnote{See Supplemental Material at [URL] for further theoretical details and additional numerical results, which includes Refs.~\cite{Weng.Jiang.2020iuo, Zheng2017, Qin2020}.} for more numerical evidence. These results suggest that when contemplating pairing mechanisms, strong Coulomb repulsion itself should be regarded as a non-negligible factor rather than simply a contributor to the spin-exchange interaction.

{\it Emergent $\mathbb{Z}_2$ Berry phases.---}In the following, we consider the $U\gg t_{\|}$ regime, in which the doubly occupied states are energetically unfavorable and can be effectively projected out. This results in the $t_{\|}\text{-}J_{\|}\text{-}J_{\perp}$ model~\cite{Wu.Lu.2023kwo, Oh2023, Su.Qu.2023, Grusdt.Bohrdt.2021, Grusdt.Bohrdt.2022, Hilker.Hirthe.2023}
\begin{equation}\label{H2}
\begin{split}
&H_{t_{\|}\text{-}J_{\|}\text{-}J_{\perp}}=-t_{\|} \sum_{\langle ij\rangle \alpha \sigma} \hat{\mathcal{P}}\left(c_{i\alpha \sigma}^{\dagger} c_{j \alpha \sigma}+\text { h.c.}\right) \hat{\mathcal{P}}\\
&- \mu \sum_{i\alpha\sigma}n_{i\alpha\sigma}+J_{\|} \sum_{\langle i j\rangle \alpha} \boldsymbol{S}_{i \alpha} \cdot \boldsymbol{S}_{j\alpha}+J_{\perp} \sum_i \boldsymbol{S}_{i 1} \cdot \boldsymbol{S}_{i 2},
\end{split}
\end{equation}
where $J_{\|}$ denotes the intralayer spin exchange coupling. Based on DFT calculations~\cite{Wang.Sun.2023, Leonov.Shilenko.2023, Kuroki.Sakakibara.2023, Werner.Christiansson.2023, Dagotto.Zhang.2023}, we find $J_{\perp}\approx 0.6 t_{\|}$, $J_{\|}\approx 2J_{\perp}/3$, and the filling is $\nu\approx 1/4$. The projector $\hat{\mathcal{P}}$ denotes the no-double-occupancy constraint $\sum_{\sigma} n_{i\alpha\sigma} \leqslant 1$. This constraint, originating from the strong Hubbard repulsion, introduces a non-trivial sign structure $\tau_C$ in the partition function
\begin{equation}\label{Zex}
    Z_{t_{\|}\text{-}J_{\|}\text{-}J_{\perp}}\equiv \operatorname{Tr} e^{-\beta H_{t_{\|}\text{-}J_{\|}\text{-}J_{\perp}}}=\sum_C \tau_C W_{t_{\|}\text{-}J_{\|}\text{-}J_{\perp}}[C],
\end{equation}
where $\beta$ is the inverse temperature, and $W_{t_{\|}\text{-}J_{\|}\text{-}J_{\perp}}[C]\geqslant 0$ denotes the non-negative weight corresponding to each closed loop $C$ of hole-spin configuration evolution. The sign factor $\tau_{C}$ is rigorously defined as follows~\cite{Note1}:
\begin{equation}\label{tau}
  \tau_{C}\equiv (-1)^{N^h_{\mathrm{ex}}} \times  (-1)^{N^h_{\downarrow}} ,
\end{equation}
in which $N^h_{\mathrm{ex}}$ represents the total number of exchanges between identical holes, like the fermionic sign structure found in doped semiconductors. The term $(-1)^{N^h_{\downarrow}}$ in Eq.~\eqref{tau} is identified as the phase-string~\cite{Weng.Sheng.1996, Ting.Weng.1997, Zaanen.Wu.2008}, in which $N^h_{\downarrow}$ accounts for the total number of mutual exchanges between holes and $\downarrow$-spins. In other words, as illustrated in \figref{fig_illu}(a), the hole hopping can accumulate a $\mathbb{Z}_2$ Berry phase with the sign determined by the orientation of the spin that is exchanged—positive for spin-up and negative for spin-down. It is crucial to emphasize that this emergent sign structure is an exclusive consequence of strong Hubbard repulsion and does not rely on specific model details. This finding is broadly applicable to other strongly correlated systems such as the single-layer $t$-$J$ model~\cite{Zaanen.Wu.2008, Lu.Weng_2023} and the Hubbard model~\cite{Weng.Zhang.2014}.

\begin{figure}
    \centering
    \includegraphics[width=0.99\linewidth]{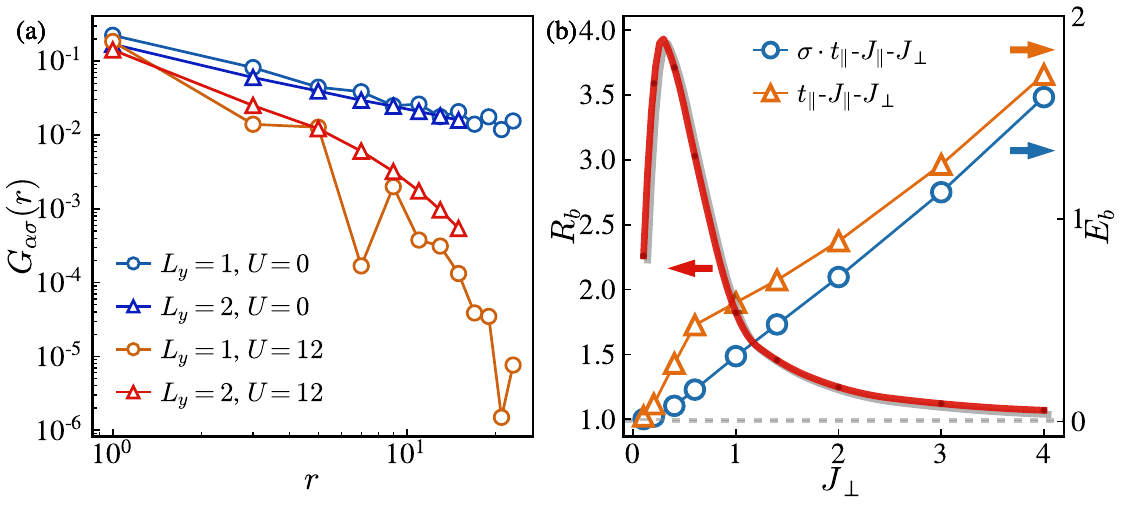}
    \caption{(a) The single-particle Green's function $G_{\alpha\sigma}(r)$ for different on-site repulsion $U$ with $J_\perp/t_\|=0.6$ for $L_y=1$ and $J_\perp/t_\|=1.4$ for $L_y=2$. (b) The binding energy $E_b$ for the $t_\|$-$J_\|$-$J_\perp$ model and $\sigma\cdot t_\|$-$J_\|$-$J_\perp$ model and the corresponding ratio $R_b$ as a function of inter-layer coupling $J_\perp$ with $J_\|/t_\|=1/3$ and system length $L_x=32$. The hole doping is fixed at $\delta=1/2$. The horizontal dashed lines mark $R_b=1$ and $E_b=0$.}
    \label{fig:green_hubbard_eb_tj}
\end{figure}

Therefore, the mobility of a single hole is significantly suppressed due to destructive interference effects among various paths, as shown in \figref{fig_illu}(a). This theoretical insight is corroborated by DMRG results, which show a rapid decay in the single-particle Green's function $G_{\alpha\sigma}(r)=\langle c_{i,\alpha\sigma}^{\dagger} c_{j+r\hat{x}, \alpha\sigma}\rangle$ in the presence of a strong Hubbard repulsion, as shown in \figref{fig:green_hubbard_eb_tj}(a)~\cite{Note1}. On the other hand, this $\mathbb{Z}_2$ frustration can be fully canceled when two holes from distinct layers pair and move together on the background of spin-singlet interlayer pairs, as depicted in \figref{fig_illu}(b). In such configurations, each exchange involving the two-hole composite and spin-singlet will consistently contribute a negative sign, which ultimately cancels out upon completing a closed loop on a bipartite lattice. This phenomenon promotes pairing between interlayer charge (and spin) degrees of freedom, steering the system towards a SC state.

To elucidate the physical implication of this unique correlation between holes and spins more explicitly, as mediated by the emergent $\mathbb{Z}_2$ Berry phase, we employ a unitary transformation~\cite{Weng.Zhu.2018tk6, Weng.Zhang.2022}
\begin{equation}\label{unitary}
    e^{i \hat{\Theta}} \equiv \exp \left(-i \sum_{i \alpha}  n_{i \alpha}^h \hat \Omega_{i\alpha}\right),
\end{equation}
on the $t_\|$-$J_\|$-$J_\perp$ model in \eqnref{H2}. Here, $n_{i \alpha}^h=1-\sum_\sigma c_{i \alpha \sigma}^{\dagger} c_{i \alpha \sigma}$ denotes the hole number at the site $i$ of layer $\alpha$. $\hat{\Omega}_{i \alpha}$ accounts for the statistical phase between the hole $n_{i \alpha}^h$ and surrounding spins within the same layer $\alpha$.

For simplicity, we first focus on the one-dimensional limit with $L_y=1$ and define $\hat \Omega_{i\alpha}^{\text{1D}}=\pi \sum_{l>i} n_{l \alpha}^{\downarrow}$ to compensate the mutual statistics between holes and spins. Thus, the original Hamiltonian is transformed into $e^{-i \hat{\Theta}} H_{ t_{\|}\text{-}J_{\|}\text{-}J_{\perp}} e^{i \hat{\Theta}}=H_{\sigma t_{\|}\text{-}J_{\|}\text{-}J_{\perp}} + H_\text{string}$. Here $H_{\sigma t_{\|}\text{-}J_{\|}\text{-}J_{\perp}}$ is defined through an additional spin-dependent sign $\sigma=\pm 1$ for $\{\uparrow,\downarrow\}$ on top of the original hopping term, i.e., $- t_{\|} c_{i \alpha \sigma}^{\dagger} c_{j \alpha\sigma} \rightarrow -\sigma t_{\|} c_{i \alpha \sigma}^{\dagger} c_{j \alpha\sigma}$, which explicit removes the sign frustration~\cite{Note1}. However, the cost needed to pay is the additional term $H_{\text{string}}=-J_{\perp}/2 \sum_i\left(S_{i 1}^{+} S_{i 2}^{-}+S_{i 1}^{-} S_{i2}^{+}\right)\left(1-\Lambda_i^h\right)$, with $\Lambda_i^h \equiv \exp \left[i \pi \sum_{l<i}\left(n_{l1}^h-n_{l2}^h\right)\right]$ characterizing the non-local phase shift effects arising from spatially separated holes on distinct chains~\cite{Note1}. Since the interlayer coupling $J_{\perp}$ implies $\langle S_{i 1}^{+} S_{i 2}^{-}+S_{i 1}^{-} S_{i2}^{+}\rangle < 0$ (denoted by $-g_0$), one can expect that two holes at the distinct chain with the coordinate $r_{\alpha=1}$ and $r_{\alpha=2}'$ will generally acquire a linear pairing potential $V_{\text{string}} \sim J_{\perp} g_0\left|r_{\alpha=1}-r_{\alpha=2}'\right|$. Such a pairing potential can be validated by comparing the binding energy in the $t_{\|}$-$J_{\|}$-$J_\perp$ and $\sigma t_{\|}$-$J_{\|}$-$J_\perp$ models through DMRG results shown in Fig.~\ref{fig:green_hubbard_eb_tj}(b), which directly demonstrate stronger binding in the $t_{\|}$-$J_{\|}$-$J_\perp$ model, particularly around the experimental relevant regime where $J_\perp/t_\|\sim0.6$~\cite{Wang.Sun.2023, Leonov.Shilenko.2023, Kuroki.Sakakibara.2023, Werner.Christiansson.2023, Dagotto.Zhang.2023}.
 
Furthermore, the above derivation can be extended to the two-dimension with $L_y\rightarrow \infty$ by setting $\hat\Omega_{i\alpha}^{\text{2D}}=\sum_{l \neq i} \theta_i(l) n_{l \alpha}^{\downarrow}$, where $\theta_{i}(l) \equiv \pm \operatorname{Im} \ln \left(z_{i}-z_{l}\right)$ with $z_i$ being the complex coordinate of site $i$, as depicted in \figref{fig_illu}(d). Similarly, the interlayer transverse coupling can also induce a strong pairing potential $V_{\text{string}} \sim J_{\perp}g_0 \left|r_{\alpha=1}-r_{\alpha=2}'\right|^2$ between two holes at distinct layers~\cite{Note1}. This additional pairing mechanism is distinguished from traditional spin-exchange pairing by its string-like long-range interaction, which means that it does not necessarily require a very strong $J_{\perp}$ to bind holes. Furthermore, while the hopping term in two dimensions under the unitary transformation \eqnref{unitary} includes additional, more complex terms compared to the one-dimensional case, these complexities do not substantially affect our preceding arguments.

{\it Effective model.---}Based on these physical implications, holes across distinct layers are inclined to form tightly bound $2e$-charge bosons (blue dashed circles in \figref{fig_illu}(c)), and spins tend to form singlet pairs across the two layers, resulting in neutral bosons with $S=0$ (red dashed circles in \figref{fig_illu}(c)). Thus, one can regard an interlayer singlet pair as a hardcore boson $ b_i=\frac{1}{\sqrt{2}}( c_{i1\uparrow} c_{i2\downarrow}-c_{i1\downarrow} c_{i2\uparrow})$ with the local particle number $\hat n_i^b=b_i^\dagger b_i\leq 1$. Then, with the standard Brillouin-Wigner perturbation theory~\cite{Note1}, the Hamiltonian in \eqnref{H2} can be reduced to an effective Hamiltonian in the low-energy subspace $\mathcal{H} = \bigotimes_i \text{span}\{|0\rangle,b_i^\dagger|0\rangle\}$, i.e.,
\begin{equation}\label{Heff}
    H_{\text{eff}}=-w\sum_{\langle ij \rangle} \left( b_i^{\dagger} b_j+\text{h.c.}\right)-V \sum_{\langle ij \rangle} \hat{n}_i^b \hat{n}_j^b-\lambda \sum_{i} \hat{n}_i^b
\end{equation}
where $w=8 t_{\|}^2/3 J_{\perp}$ denotes the effective hopping, $V=J_{\|}^2/8 J_{\perp}$ denotes a weak neighboring attraction with $V \ll w$ since $t_{\|}\gg J_{\|}$, and $\lambda = 3 J_{\perp}/4+2 \mu$ is the chemical potential associated with the filling $\nu$. 

The effective model $H_\text{eff}$ in \eqnref{Heff} describes a single-layer hard-core Bose system with a weak nearest-neighbor attractive interaction, as shown in \figref{fig_illu}(c). In the one-dimensional scenario, $H_\text{eff}$ simplifies to a single chain and can be analytically solved using the bosonization method after applying the Jordan-Wigner transformation~\cite{Note1}. This results in a Luther-Emery liquid, characterized by an SC exponent $K_{\text{SC}}<1$. In the two-dimensional scenario, one can perform a self-consistent mean-field calculation with the order parameter $\langle b_i \rangle=\sqrt{\rho_s}$~\cite{Note1}. Our findings indicate that the $b$-bosons will enter a superfluid phase with $\rho_s \neq 0$ at zero temperature unless $2\nu$ is an integer. This phase corresponds to the SC phase in the original fermionic system, marked by the condensation of Cooper pairs. At finite temperatures, the coherence of singlet pairs would be disrupted by the thermal fluctuation and then undergo a Kosterlitz-Thouless (KT) transition at the critical temperature $T_c$. As a result, the SC phase transition temperature can be estimated by the KT temperature~\cite{RevModPhys.51.659} $T_c =\frac{8 \pi t_{\|}^2}{ 3 J_{\perp}}\rho_s$,
of which the filling dependence is shown in \figref{fig_illu}(e). Our results indicate that $T_c$ peaks at $\nu=1/4$, aligning with the actual situation of the $d_{x^2-y^2}$ orbitals in LNO. In contrast, $T_c$ is drastically suppressed at $\nu=1/2$, associated with the undoped featureless Mott insulator, also known as a symmetric mass generation (SMG) insulator~\cite{You.Wang.2022, You.Hou.2022, You.Lu.2023a, You.Lu.2023b, You.Lu.2023c}.

{\it Discussion.---}In this work, the pairing mechanism in the high-$T_c$ SC material La$_3$Ni$_2$O$_7$ has been explored, in which a significant role of the Coulomb repulsion $U$ among $d$-orbital electrons has been revealed. Although a non-zero interlayer superexchange coupling $J_\perp$ is essential for the pairing, the binding energy is found to be strongly enhanced with $U$ tuned into the Mott regime as opposed to $U=0$. In the former regime (large $U$), the propagation of the single electron (hole) becomes severely frustrated due to the accumulation of a $\mathbb{Z}_2$ Berry phase (phase-string). Such frustration can be completely canceled out when two holes are paired up with the help of $J_\perp$ such that the strong suppression of the kinetic energy can be effectively released. The recent development of flat-band superconductors can be regarded as a single-particle version of this picture. In that case, the kinetic energy is quenched due to the lattice interference pattern, but it can be restored after forming Cooper pairs~\cite{2022NatRP...4..528T, Peotta2015SuperfluidityIT, PhysRevLett.132.026002}. By contrast, by turning off this Berry phase artificially, the binding is also suppressed even in the large $U$ case. It means that the pairing in the Mott regime is indeed substantially driven up by saving the kinetic energy~\cite{Weng.Zhao.2022, Weng.Chen.2018}, with the magnitude of about $10$-$20$ times larger than that in the $U=0$ case around $J_\perp/t_\|\sim1$. The latter regime (small $U$) has been studied in other approaches~\cite{You.Lu.2023}. Note that though the latter mechanism may also yield a similar effective Hamiltonian as in \eqnref{Heff} in the BEC limit~\cite{You.Lu.2023}, it generally requires a much larger $J_\perp$ exceeding the hopping integral $t_{\|}$. 

Moreover, at the high-temperature phase, the spin-singlet pairs are destroyed, and hence the strong $\mathbb{Z}_2$ Berry phase frustration that can be experienced by both single hole and hole pairs, resulting in a loss of coherence of the charge degrees of freedom. The scattering between holes and random phase (flux) can naturally give rise to a linear-$T$ dependence of the electric resistivity~\cite{Weng.Gu.2007, chenzhang2024}, which is consistent with the experimental measurement in LNO~\cite{Yuan.Zhang.2023, Wen.Liu.2023,Wang.Sun.2023}.

\begin{acknowledgments}
{\it Acknowledgments.---}We acknowledge stimulating discussions with Fan Yang, Guang-Ming Zhang, Shuai Chen, Ji-Si Xu, and Zhao-Yi Zeng. We are especially grateful to Fan Yang for his valuable suggestions on material modeling. J.-X.Z., H.-K.Z, Z.-Y.W. are supported by MOST of China (Grant No. 2021YFA1402101) and NSF of China (Grant No. 12347107); Y.-Z.Y. is supported by the National Science Foundation Grant No. DMR-2238360.
\end{acknowledgments}

%


\clearpage
\newpage
\widetext

\begin{center}
\textbf{\large Supplementary Material for \\``Strong Pairing Originated from an Emergent $\mathbb{Z}_2$ Berry Phase in La$_3$Ni$_2$O$_7$''}
\end{center}


\renewcommand\thefigure{\thesection S\arabic{figure}}
\renewcommand\theequation{\thesection S\arabic{equation}}

\setcounter{figure}{0} 
\setcounter{equation}{0}

In this supplementary material, we offer additional analytical and numerical results to reinforce the conclusions in the main text. In Section I, we present a rigorous derivation of the sign structure for both the $t_{\|}$-$J_{\|}$-$J$ model and the $\sigma t_{\|}$-$J_{\|}$-$J$ model. Section II offers a comprehensive derivation of the effective attraction interaction, denoted as $V_{\text{string}}$, for both one-dimensional and two-dimensional scenarios. In Section III, we delve into the derivation of the effective model $H_{\text{eff}}$ and subsequently explore its physical implications in both one-dimensional and two-dimensional cases. In Section IV, we show more numerical evidences on the enhancement of pairing strength from the on-site repulsion.

\section{I. The Exact Sign Structure of the $t_{\|}\text{-}J_{\|} \text{-}J_{\perp}$ Hamiltonian}
In this section, we provide rigorous proof for the partition function of the $t_{\|}$-$J_{\|}$-$J$ model and elaborate on the sign structure. We initiate our discussion with the slave-fermion formalism. In this context, the electron operator is represented as $c_{i\alpha\sigma}=f_{i\alpha}^{\dagger} b_{i\alpha\sigma}$, where $f_{i\alpha}^{\dagger}$ refers to the fermionic holon operator and $b_{i\alpha\sigma}$ to the bosonic spinon operator. These relations are governed by the constraint $f_{i\alpha}^{\dagger} f_{i\alpha}+\sum_{\sigma} b_{i\alpha\sigma}^{\dagger} b_{i \alpha\sigma}=1$. To shed light on the sign structure inherent in this model, we incorporate the Marshall sign into the $S_z$-spin representation. This is achieved through the substitution $b_{i \alpha\sigma}\rightarrow (-\sigma)^{i+\alpha} b_{i \alpha\sigma}$, which leads to
\begin{equation}\label{frac}
	c_{i \alpha\sigma}=(-\sigma)^{i+\alpha} f_i^{\dagger} b_{i \alpha\sigma}.
\end{equation}
Consequently, the $\sigma t_{\|}\text{-}J_{\|} \text{-}J_{\perp}$ model is transformed as follows: 
\begin{equation}
  H_{t_{\|}\text{-}J_{\|}\text{-}J_{\perp}}=-t_{\|}\left(P_{o \uparrow}^{\|}-P_{o \downarrow}^{\|}\right)-\frac{J_{\|}}{2} \left(Q^{\|}+P_{\uparrow \downarrow}^{\parallel}\right)-\frac{J_{\perp}}{2} \left(Q^{\perp}+P_{\uparrow \downarrow}^{\perp}\right),
\end{equation}
where
\begin{eqnarray}
  P_{o \uparrow}^{\parallel}&=&\sum_{\langle ij\rangle \alpha} b_{i \alpha\uparrow}^{\dagger} b_{j \alpha\uparrow} f_{j}^{\dagger} f_{i}+\text {h.c.}, \;\;\;\;\;\;\;\;\;\;\;\;
  P_{o \downarrow}^{\parallel}=\sum_{\langle ij\rangle \alpha} b_{i \alpha\downarrow}^{\dagger} b_{j \alpha\downarrow} f_{j}^{\dagger} f_{i}+\text {h.c.}, \\
  P^{\parallel}_{\uparrow \downarrow}&=&\sum_{\langle i j\rangle \alpha} b_{i\alpha \uparrow}^{\dagger} b_{j\alpha \downarrow}^{\dagger} b_{i\alpha \downarrow} b_{j\alpha \uparrow}+\text {h.c.}, \;\;\;\;\;\;
  P^{\perp}_{\uparrow \downarrow}=\sum_{i} b_{i1 \uparrow}^{\dagger} b_{i2 \downarrow}^{\dagger} b_{i1 \downarrow} b_{i2 \uparrow}+\text {h.c.}, \\  Q^{\parallel}&=&\sum_{\langle ij\rangle \alpha}\left(n_{i \uparrow} n_{j \downarrow}+n_{i \downarrow} n_{j \uparrow}\right), \;\;\;\;\;\;\;\;\;\;\;\;\;\;
Q^{\perp}=\sum_{i}\left(n_{i1 \uparrow} n_{i2 \downarrow}+n_{i1 \downarrow} n_{i2 \uparrow}\right)
\end{eqnarray}
The operators $P_{o \sigma}^{\|}$, $P_{\uparrow \downarrow}^{\|}$, and $P_{\uparrow \downarrow}^{\perp}$ represent the intralayer hole-spin nearest-neighbor (NN) exchange, intralayer NN spin superexchange, and interlayer NN spin superexchange, respectively. The terms $Q^{\|}$ and $Q^{\perp}$ correspond to potential interactions between NN spins. By utilizing the high-temperature series expansion, we derive the partition function up to all orders~\cite{Zaanen.Wu.2008} 
\begin{equation}\label{de}
\begin{aligned}
&Z_{t_{\|}\text{-}J_{\|}\text{-}J_{\perp}}=\operatorname{Tr} e^{-\beta H_{t_{\|}\text{-}J_{\|}\text{-}J_{\perp}}}=\operatorname{Tr} \sum_{n=0}^{\infty} \frac{\beta^{n}}{n !}\left(-H_{t_{\|}\text{-}J_{\|}\text{-}J_{\perp}}\right)^{n} \\
&=\sum_{n=0}^{\infty}  \frac{(J_{\perp} \beta / 2)^{n}}{n !} \operatorname{Tr}\left[\sum \ldots\left(\frac{2 t_{\|}}{J_{\perp}} P^{\|}_{o \uparrow}\right) \ldots \left(\frac{J_{\|}}{J_{\perp}} Q^{\|}\right) \ldots P^{\perp}_{\uparrow \downarrow} \ldots\left(-\frac{2 t_{\|}}{J_{\perp}} P_{o \downarrow}^{\|}\right) \ldots\left(\frac{J_{\|}}{J_{\perp}} P_{\uparrow \downarrow}^{\|}\right) \ldots Q^{\perp} \ldots\right]_{n} \\
&=\sum_{n=0}^{\infty}(-1)^{N_{ \downarrow}^h} \frac{(J_{\perp} \beta / 2)^{n}}{n !} \operatorname{Tr}\left[\sum \ldots\left(\frac{2 t_{\|}}{J_{\perp}} P^{\|}_{o \uparrow}\right) \ldots \left(\frac{J_{\|}}{J_{\perp}} Q^{\|}\right) \ldots P^{\perp}_{\uparrow \downarrow} \ldots\left(\frac{2 t_{\|}}{J_{\perp}} P_{o \downarrow}^{\|}\right) \ldots\left(\frac{J_{\|}}{J_{\perp}} P_{\uparrow \downarrow}^{\|}\right) \ldots Q^{\perp} \ldots\right]_{n}
\end{aligned}
\end{equation}
with the underlying assumption that the NN hopping integral remains positive, signified by $t_{\|}>0$. The notation $[\sum \ldots]_n$ encompasses the summation over all $n$-block production. Owing to the trace, both starting and ending configurations of holes and spins must coincide, ensuring that all contributions to $Z_{t_{\|}\text{-}J_{\|}\text{-}J_{\perp}}$ are delineated by closed loops of holes and spins. Within this framework, $N_{\downarrow}^h$ quantifies the exchanges between down-spins and holes. Then, we insert a complete Ising basis with holes, given by $\sum_{\phi\left\{l_{h}\right\}}\left|\phi ;\left\{l_{h}\right\}\right\rangle\left\langle\phi ;\left\{l_{h}\right\}\right|=1$, between operators inside the trace.  Here, $\phi$ specifies the spin configuration, while $\left\{l_{h}\right\}$ indicates the positions of the holes. Therefore, the partition function can be written as a compact expression
\begin{equation}
  Z_{t_{\|}\text{-}J_{\|}\text{-}J_{\perp}}=\sum_{C} \tau_{C} W_{t_{\|}\text{-}J_{\|}\text{-}J_{\perp}}[C],
\end{equation}
with the relevant sign information is encapsulated by
\begin{equation}\label{tautotal}
  \tau_{C}^{t_{\|}\text{-}J_{\|}\text{-}J_{\perp}}\equiv (-1)^{N_{\mathrm{ex}}^h} \times(-1)^{N_{\downarrow}^h},
\end{equation}
which aligns with the main text. In this context, $N_{\mathrm{ex}}^h$ represents the number of hole exchanges arising from the fermionic statistics of the holon $f$. This sign structure is comprehensively defined across varying doping levels, temperatures, and finite sizes for the $t_{\|}\text{-}J_{\|}\text{-}J_{\perp}$ model. Additionally, the non-negative weight $W[C]$ for a closed loop $C$ emerges from a sequence of positive elements within the trace of \eqnref{de}.

Furthermore, by introducing the $\sigma t_{|}\text{-}J_{|} \text{-}J_{\perp}$ model, where the intrinsic spin exchange term remains unchanged, but the hopping term is substituted by
\begin{equation}
  H_{\sigma t_{\|}}=- t_{\|} \sum_{\langle i j\rangle} \sigma c_{i \sigma}^{\dagger} c_{j \sigma}+\text {h.c.},
\end{equation}
in which an additional spin-dependent sign, $\sigma$, is integrated into the NN hopping term, cancelling the ``$-$'' sign preceding the $P_{o \downarrow}$. As a result, the $\sigma t_{|}\text{-}J_{|} \text{-}J_{\perp}$ model, under the representation of Eq. \eqref{frac}, can be rewritten as
\begin{equation}
  H_{\sigma  t_{\|}\text{-}J_{\|} \text{-}J_{\perp}} = -t_{\|} \left(P_{o \uparrow}^{\|} + P_{o \downarrow}^{\|}\right)-\frac{J_{\|}}{2} \left(Q^{\|} + P_{\uparrow \downarrow}^{\|}\right) - \frac{J_{\perp}}{2} \left(Q^{\perp} + P_{\uparrow \downarrow}^{\perp}\right),
\end{equation}
with the partition function under the high-temperature series expansion expressed as
\begin{eqnarray}
  Z_{\sigma  t_{\|}\text{-}J_{\|} \text{-}J_{\perp}} =\sum_{n=0}^{\infty} \frac{(J_{\perp} \beta / 2)^{n}}{n !} \operatorname{Tr}\left[\sum \ldots\left(\frac{2 t_{\|}}{J_{\perp}} P^{\|}_{o \uparrow}\right) \ldots \left(\frac{J_{\|}}{J_{\perp}} Q^{\|}\right) \ldots P^{\perp}_{\uparrow \downarrow} \ldots\left(\frac{2 t_{\|}}{J_{\perp}} P_{o \downarrow}^{\|}\right) \ldots\left(\frac{J_{\|}}{J_{\perp}} P_{\uparrow \downarrow}^{\|}\right) \ldots Q^{\perp} \ldots\right]_{n}
\end{eqnarray}
As the result, the sign structure for $\sigma  t_{\|}\text{-}J_{\|} \text{-}J_{\perp}$ model is given by
\begin{equation}\label{taustJ}
  \tau_{C}^{\sigma  t_{\|}\text{-}J_{\|} \text{-}J_{\perp}}= (-1)^{N_{\mathrm{ex}}^h} ,
\end{equation}
in which the nontrivial phase string sign $(-1)^{N_{\downarrow}^h}$ vanishes.

In the end, we further clarify the role played by the strong Hubbard repulsion itself in this model. As the hopping integral between the two layers vanishes in the Hamiltonian in Eq.~(\textcolor{blue}{1}) of the main text, projecting our the double occupied states is actually the same as the conventional perturbation process from the single-layer Hubbard model to the $t$-$J$ model. Note that the $J_\perp$ term only involves single occupied states and hence it remains in the same form after the projection. In other words, except for the interlayer superexchange interaction especially in this bilayer lattice, the strong Hubbard repulsion plays the same role as that in the conventional single-layer Mott insulator: opening the Mott gap and hence effectively constraining the Hilbert space by the non-double occupancy projection.

\begin{figure}
    \centering
    \includegraphics[width=0.7\linewidth]{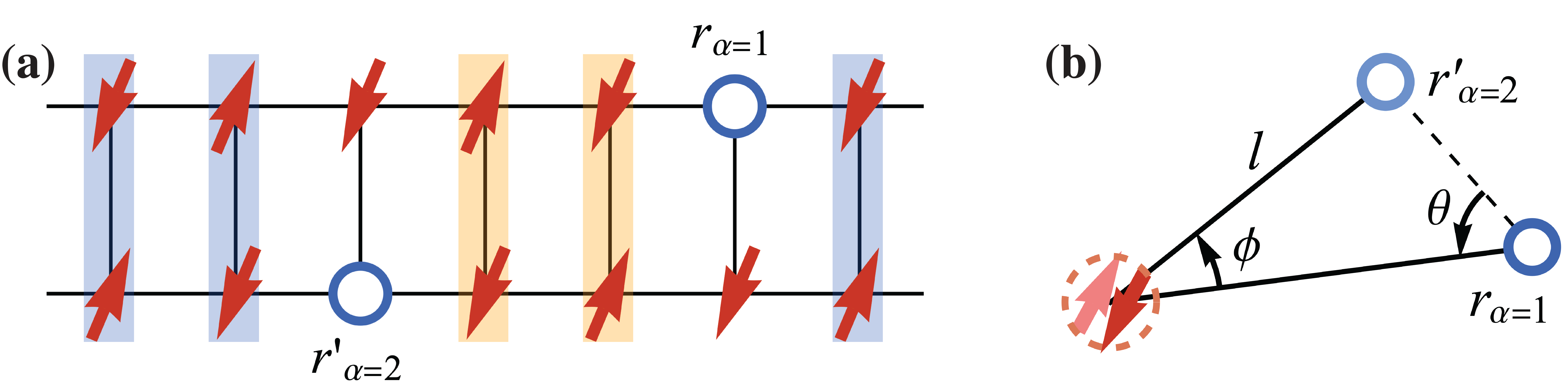}
    \caption{(a) Illustration of a certain configuration for holes (marked by blue) and spins (marked by red). $H_{\text{string}}$ vanishes at the blue rung, while it remains finite at the orange rung. (b) Sketch of positions of holes and spins on a two-layer system, with lighter (darker) color labeling the objects on the first(second) layer.}
    \label{fig_app1}
\end{figure}

\section{II. Effective Attraction Interaction $V_{\text{string}}$}
In this section, more derivation details about the effective attraction interaction $V_{\text{string}}$ in the main text will be discussed for both the one-dimensional and two-dimensional cases.

\subsection{II.A. One Dimensional Case}
In the one-dimensional context, upon employing a unitary transformation $e^{i \hat{\Theta}} \equiv \exp \left(-i \sum_{i \alpha} n_{i \alpha}^h \hat{\Omega}_{i \alpha}\right)$ with $\hat{\Omega}_{i \alpha}^{1 \mathrm{D}}=\pi \sum_{l>i} n_{l \alpha}^{\downarrow}$, the hopping term in $t_{\|}$-$J_{\|}$-$J_{\perp}$ model is given by~\cite{Weng.Jiang.2020iuo}
\begin{eqnarray}\label{Htch}
    e^{-i \hat{\Theta}} c_{i \alpha \sigma}^{\dagger} c_{i+1 \alpha \sigma} e^{i \hat{\Theta}}&=&\exp \left[-i \sum_{j, \alpha} n_{j \alpha}^h \pi \sum_{l>j} n_{l \alpha}^{\downarrow}\right] c_{i \alpha \sigma}^{\dagger} c_{i+1 \alpha \sigma} \exp \left[i \sum_{k, \alpha} n_{k \alpha}^h \pi \sum_{l>k} n_{l\alpha}^{\downarrow}\right]\\
	&=&\exp \left[-i \sum_{j=i,i+1} \sum_{\alpha} n_{j \alpha}^h \pi \sum_{l>j} n_{l \alpha}^{\downarrow}\right] c_{i \alpha \sigma}^{\dagger} c_{i+1 \alpha \sigma} \exp \left[i \sum_{k\neq i,i+1}\sum_{\alpha} n_{k \alpha}^h \pi \sum_{l>k} n_{l \alpha}^{\downarrow}\right] \\
	&=&c_{i \alpha \sigma}^{\dagger} c_{i+1 \alpha \sigma} \exp \left[-i \pi n_{i+1 \alpha}^{\downarrow}\right] \\
	&=&\sigma c_{i \alpha, \sigma}^{\dagger} c_{i+1 \alpha \sigma}.
  \end{eqnarray}
Similarly, the spin exchange interaction transforms as:
  \begin{eqnarray}\label{Hzch}
	e^{-i \hat{\Theta}} S_{i 1}^z S_{i 2}^z e^{i \hat{\Theta}}=\exp \left[-i \sum_{j, \alpha} n_{j \alpha}^h \pi \sum_{l>j} n_{l, \alpha}^{\downarrow}\right] c_{i 1 \uparrow}^{\dagger} c_{i 1 \downarrow} c_{i 2 \downarrow}^{\dagger} c_{i 2 \uparrow} \exp \left[i \sum_{k, \alpha} n_{k \alpha}^h \pi \sum_{l>k} n_{l \alpha}^{\downarrow}\right]=S_{i 1}^z S_{i 2}^z,
  \end{eqnarray}
as well as:
  \begin{eqnarray}\label{Hpmch}
	e^{-i \hat{\Theta}} S_{i 1}^+ S_{i 2}^- e^{i \hat{\Theta}}&=&\exp \left[-i \sum_{j, \alpha} n_{j \alpha}^h \pi \sum_{l>j} n_{l \alpha}^{\downarrow}\right] c_{i 1 \uparrow}^{\dagger} c_{i 1 \downarrow} c_{i 2 \downarrow}^{\dagger} c_{i 2 \uparrow} \exp \left[i \sum_{k, \alpha} n_{k \alpha}^h \pi \sum_{l>k} n_{l \alpha}^{\downarrow}\right]\\
	&=&\exp \left[-i \sum_{j<i, \alpha} n_{j \alpha}^h \pi \sum_{l>j} n_{l \alpha}^{\downarrow}\right] c_{i 1 \uparrow}^{\dagger} c_{i 1 \downarrow} c_{i 2 \downarrow}^{\dagger} c_{i 2 \uparrow} \exp \left[i \sum_{k<i, \alpha} n_{k \alpha}^h \pi \sum_{\downarrow>k} n_{l \alpha}^{\perp}\right]\\
	&=&S_{i 1}^{+} S_{i 2}^{-} \exp \left[i \pi \sum_{l<i}\left(n_{l 1}^h-n_{l 2}^h\right)\right].
  \end{eqnarray}
As the result, combining \eqnref{Htch}, \eqnref{Hzch} and \eqnref{Hpmch}, the total Hamltonian $t_{\|}$-$J_{\|}$-$J_{\perp}$ can be expressed as:
\begin{equation}
	e^{-i \hat{\Theta}} H_{t_{\|}-J_{\|}-J_{\perp}} e^{i \hat{\Theta}}=H_{\sigma t_{\|}-J_{\|}-J_{\perp}}+H_{\text {string}}.
\end{equation}
Here, $H_{\sigma t_{\|}-J_{\|}-J_{\perp}}$ and $H_{\text {string }}$ are defined as:
\begin{eqnarray}
	H_{\sigma t_{\|}-J_{\|}-J_{\perp}}&=&-t_{\|} \sum_{\langle i j\rangle \alpha \sigma}\left(\sigma c_{i \alpha \sigma}^{\dagger} c_{j \alpha \sigma}+\text { h.c. }\right)-\mu \sum_{i \alpha \sigma} n_{i \alpha \sigma}+J_{\|} \sum_{\langle i j\rangle \alpha} \boldsymbol{S}_{i \alpha} \cdot \boldsymbol{S}_{j \alpha}+J_{\perp} \sum_i \boldsymbol{S}_{i 1} \cdot \boldsymbol{S}_{i 2}\label{HstJ}\\
	H_{\text {string }}&=& \frac{J_{\perp}}{2} \sum_i\left(S_{i 1}^{+} S_{i 2}^{-}+S_{i 1}^{-} S_{i 2}^{+}\right)\left(\Lambda_i^h-1\right)\label{Hstring}
\end{eqnarray}
with $\Lambda_i^h=\exp \left[i \pi \sum_{l<i}\left(n_{l 1}^h-n_{l 2}^h\right)\right]$. Here, \eqnref{HstJ} is the $\sigma t_{\|^{-}} J_{\|^{-}} J_{\perp}$ model with only the fermionic statistic sign, as depicted in \eqnref{taustJ}. Furthermore, given the expectation value of the transverse spin at each rung by $\langle S_{i 1}^{+} S_{i 2}^{-}+S_{i 1}^{-} S_{i2}^{+}\rangle < 0$ (denoted by $ - g_0$),  it becomes evident that \eqnref{Hstring} vanishes when hole pairs across the rung exist on a background where spins also establish interchain singlet pairing. Nonetheless, when these hole pairs are broken and the two holes are separated by some distance, every rung with spin-singlet pairs located between these holes will contribute $J_{\perp} g_0 / 2$ (highlighted by orange shadows in \figref{fig_app1}(a)). This results in an energy cost that depends linearly: 
\begin{equation}
	V_{\text {string }} \sim J_{\perp} g_0\left|r_{\alpha=1}-r_{\alpha=2}^{\prime}\right|,
\end{equation}
where $r_{\alpha=1}$ and $r_{\alpha=2}^{\prime}$ denote the coordinates of the holes on chain 1 and chain 2, respectively.

\subsection{II.B. Two Dimensional Case}
In the two dimensional case, the unitary transformation can be generalized by setting $\hat{\Omega}_{i \alpha}^{2 \mathrm{D}}=\sum_{l \neq i} \theta_i(l) n_{l \alpha}^{\downarrow}$. With this transformation, the spin exchange term evolves as:
\begin{eqnarray}
	e^{-i \hat{\Theta}} S_{i 1}^{+} S_{i 2}^{-}  e^{i \hat{\Theta}}=S_{i 1}^{+} S_{i 2}^{-} \exp \left[i \sum_{l \neq i} \theta_i(l)\left(n_{l 1}^h-n_{l 2}^h\right)\right].
\end{eqnarray}
When spin singlets form across the two layers, i.e., $\left\langle S_{i 1}^{+} S_{i 2}^{-} \right\rangle <0$, the ensuing energy cost from this term can be expressed as:
\begin{eqnarray}\label{Vstring2D}
	V_{\text{string}} &=& \left \langle \frac{J_{\perp}}{2} \sum_i e^{-i \hat{\Theta}}\left( S_{i 1}^{+} S_{i 2}^{-}  + S_{i 1}^{-} S_{i 2}^{+}   \right)e^{i \hat{\Theta}}-\frac{J_{\perp}}{2} \sum_i \left( S_{i 1}^{+} S_{i 2}^{-}  + S_{i 1}^{-} S_{i 2}^{+}   \right)\right \rangle\\
	&=& -J_{\perp} g_0 \sum_i \cos \left[i \sum_{l \neq i} \theta_i(l)\left(n_{l 1}^h-n_{l 2}^h\right)\right]+\sum_i Jg_0\\
	&\approx&  \frac{J_{\perp} g_0}{2} \int_0^{\infty} r d r \int_0^{2 \pi} d \theta \phi^2 + E_{\text{core}}\\
	&\varpropto&J_{\perp} g_0 \ln R  \left|r_{\alpha=1}-r_{\alpha=2}^{\prime}\right|^2+E_{\text {core }},
\end{eqnarray}
where $R$ is the sample size, and $E_{\text {core }}$ represents the ultraviolet energy in proximity to the hole pairs. In the last line of \eqnref{Vstring2D}, the relation $\phi \simeq |r_{\alpha=1}-r_{\alpha=2}^{\prime}| \sin \theta/l $ is used, and the definition of $\phi$, $l$ and $\theta$ is illustrated in \figref{fig_app1}(b).  It is noteworthy that the string tension $\ln R$ in \eqnref{Vstring2D} approaches infinity as we tend towards the thermodynamic limit. However, the term $g_0 \ln R$ aims to remain finite, thereby avoiding energy divergence.

\section{III. Effective Model $H_{\text{eff}} $}
In this section, more derivation about the hard-core bosonic effective Hamiltonian $H_{\text{eff}}$ will be given, and then we will discuss its physical implication using the one-dimensional Abelian bosonization approach and the two-dimensional mean-field theory, respectively.

From our preceding discussions, the local Hilbert space at the $i$th rung of the original double-layer system can be divided into two subspaces with distinct energy scales:
\begin{itemize}
	\item low-energy sector $\mathcal{V}_0$:
	\begin{eqnarray}
		|00\rangle_i&:& E=0 \\
		\frac{1}{\sqrt{2}}(|\uparrow \downarrow\rangle_i-|\downarrow \uparrow\rangle_i)&:& E=-\frac{3}{4} J_{\perp}
	  \end{eqnarray}
	\item high-energy sector $\mathcal{V}_1$:
	\begin{eqnarray}
		|\downarrow \downarrow\rangle_i,\;\;\;\;|\uparrow \uparrow\rangle_i, \;\;\;\;\frac{1}{\sqrt{2}}(|\uparrow \downarrow\rangle_i+|\downarrow \uparrow\rangle_i)&:& E=\frac{1}{4} J_{\perp} \\
		|\downarrow 0\rangle_i,\;\;\;\;|\uparrow 0\rangle_i,\;\;\;\;|0\downarrow\rangle_i,\;\;\;\;|0\uparrow\rangle_i&:& E=V_{\text {string }}
	  \end{eqnarray}
\end{itemize}
Then, we can define projective operators $P$ onto the low-energy sector $\mathcal{V}_0$ as
\begin{equation}
	P=\prod_i \sum_{\alpha \in V_0}|\alpha\rangle\langle\alpha|,
\end{equation}
and  projective operators $Q$ onto the high-energy sector $\mathcal{V}_1$;
\begin{equation}
	Q=1-P.
\end{equation}
Applying the Brillouin-Wigner perturbation theory, the effective Hamiltonian is given by: $H_{\mathrm{eff}}(E)=P H P-P H Q(Q H Q-E)^{-1} Q H P$. At the zeroth order, it becomes
\begin{equation}\label{0th}
	P H_{\sigma t_{\|}-J_{\|}-J_{\perp}} P =-\frac{3}{4} J_{\perp} \sum_i n_i^b-2 \mu \sum_i n_i^b,
\end{equation}
where $n_i^b=b_i^\dagger b_i$ denotes the local number operator for hard-core boson, with the correspondence $b_i=\frac{1}{\sqrt{2}}\left(c_{i 1 \uparrow} c_{i 2 \downarrow}-c_{i 1 \downarrow} c_{i 2 \uparrow}\right)$. For the hopping term, the second-order virtual process is expressed as
\begin{equation}\label{2nd1}
	P H_{t_{\|}} Q(E-Q H Q)^{-1} Q H_{t_{\|}} P=-\frac{8 t_{\|}^2}{3 J_{\perp} + 4V_{\text{string}}}\left(b_i^{\dagger} b_j+h . c .\right).
\end{equation}
Similarly, for the intralayer spin exchange interactions, the second-order virtual process is given by
\begin{equation}\label{2nd2}
	P H_{J_{\|}} Q(E-Q H Q)^{-1} Q H_{J_{\|}} P=-\frac{J_{\|}^2}{8 J_{\perp}} n_i^b n_j^b.
\end{equation}
As the result, combining Eqs.~\eqref{0th}, \eqref{2nd1} and \eqref{2nd2}, the resulting effective Hamltonian is given by
\begin{equation}\label{Heffapp}
	H_{\mathrm{eff}}=\sum_i-w\left(b_i^{\dagger} b_j+\text { h.c. }\right)-V \hat{n}_i^b \hat{n}_j^b-\lambda \hat{n}_i^b,
\end{equation}
which is consistent with the main text. Here, $w\approx 8 t_{\|}^2/3 J_{\perp}$ denotes effective hopping term for boson, $V=J_{\|}^2/8 J_{\perp}$ denotes a weak neighbouring attractive interaction and $\lambda =  3 J_{\perp}/4+2 \mu$ is the chemical potential associated with the fillings $\nu$. Note that $V_{\text{string}}$ here becomes negligible when the separation distance is very short (only a lattice constant in this case).

\subsection{III.A. One Dimensional Case: Bosonization}
For the one-dimensional scenario, the effective Hamiltonian can be expressed as
\begin{equation}\label{Hpsi}
	H[\psi]=-w \sum_{i}\left(\psi_i^{\dagger} \psi_{i+1}+\psi_{i+1}^{\dagger} \psi_i\right)-V \sum_{ i} n_i^b n_{i+1}^b-\lambda \sum_i n_i^b
\end{equation}
under the Jordan-Wigner transformation
\begin{equation}
	\psi_i=b_i \exp \left(-i \pi \sum_{l>i} n_i\right).
\end{equation}
Upon employing standard Abelian bosonization, the relationship between the fermionic and bosonic fields is
\begin{equation}
	\psi_i=\frac{\eta}{\sqrt{2 \pi a}} e^{i k_b x} e^{-i\left(\phi_b-\theta_b\right)}+\frac{\bar{\eta}}{\sqrt{2 \pi a}} e^{-i k_b x} e^{-i\left(-\phi_b-\theta_b\right)}.
\end{equation}
Consequently, the fermionic Hamiltonian, denoted \eqnref{Hpsi}, transforms to
\begin{equation}
	H[\phi, \theta]=\frac{u_b}{2 \pi} \int d x\left\{K_b\left[\partial_x \theta_b(x)\right]^2+\frac{1}{K_b}\left[\partial_x \phi_b(x)\right]^2\right\}-\frac{U}{\pi} \partial_x \phi_b,
\end{equation}
which illustrates the low-energy fluctuations of the pairing field, \(\phi\), with the stiffness constant
\begin{eqnarray}\label{Kb}
K_b= 1+V \frac{2 \sin ^2 k_F }{\pi u_b}>1,
\end{eqnarray}
where $k_F=2\pi \nu$ signifies the effective Fermi momentum. Additionally, the  density-density correlation with the hole number operator, denoted by \(n^h_i = 1-\sum_{\alpha\sigma}n_{i\alpha\sigma}/2\), is:
\begin{equation}\label{nnc}
	\left\langle n^h_i n^h_{i+r}\right\rangle=\left\langle n_i^b n_{i+r}^b\right\rangle \sim \frac{2}{(2 \pi)^2}|r|^{-2 K_b} \cos \left(2 k_F r\right),
\end{equation}
where the expression for the density operator is given by $n_x^b=-\frac{1}{\pi} \partial_x \phi_b+\frac{1}{2 \pi}\left[e^{i 2 k_b x} e^{-2 i \phi_b}+\text { h.c. }\right]$. As a result, from \eqnref{nnc}, we deduce that the charge density exponent is $K_C=2 K_b$. Moreover, the singlet superconducting pairing correlation, $\left\langle b_i b_{i+r}^{\dagger}\right\rangle$ in the fermionic representation is:
\begin{equation}
	\left\langle b_i b_{i+r}^{\dagger}\right\rangle=\left\langle\exp \left(-i \pi \sum_{l>i} \hat{n}_l\right) \psi_i \psi_{i+r}^{\dagger} \exp \left(+i \pi \sum_{l>i+r} \hat{n}_l\right)\right\rangle\sim e^{i k_F r}|r|^{-(1 / 2) K_b^{-1}},
\end{equation}
so that $K_{SC}=\frac{1}{2} K_b^{-1}$. A key observation is that $K_C K_{SC}=1$, which characterizes a Luther-Emery liquid. Notably, for $V>0$ indicating an attractive interaction, $K_b>1$ from \eqnref{Kb} implies a heightened SC instability with \(K_{SC}<1\).

\subsection{III.B. Two Dimensional Case: Mean-Field Theory}
In the two-dimensional case, the Hilbert space for distinct sites can be decoupled using the mean field ansatz, \(\left\langle b_i\right\rangle=\sqrt{\rho_s}\). The resulting mean-field (MF) Hamiltonian becomes:
\begin{equation}
	H_{\text{MF}}= \sum_i\left[-s \left(b_i^{\dagger}+ b_i \right )-v n_i^b +\epsilon \right]
\end{equation}
with
\begin{equation}
	s= \frac{32 }{3 J_{\perp}} t_{\|}^2\sqrt{\rho_s}, \;\;\;\;\;v=\frac{J_{\|}^2}{2 J_{\perp}} \rho_s+\frac{3}{4} J_{\perp}+2 \mu, \;\;\;\;\;\;\;\epsilon= \frac{32 }{3 J_{\perp}} t_{\|}^2 \rho_s+ \frac{J_{\|}^2}{4 J_{\perp}} \rho_s^2.
\end{equation}
By diagonalizing this two-dimensional on-site Hilbert space, we acquire the local ground state \(|\text{GS}\rangle_i\) and its corresponding energy \(E_{\text{GS}}(\rho_s)\). Both \(\rho_s\) and \(\mu\) can be determined by minimizing \(E_{\text{GS}}(\rho_s)\) with respect to $\rho_s$, and preserving the filling condition: \(\langle \text{GS}| n_i^b | \text{GS} \rangle=2\nu\).

\subsection{IV. Additional numerical results and technical details}
\begin{figure}
    \centering
    \includegraphics[width=0.7\linewidth]{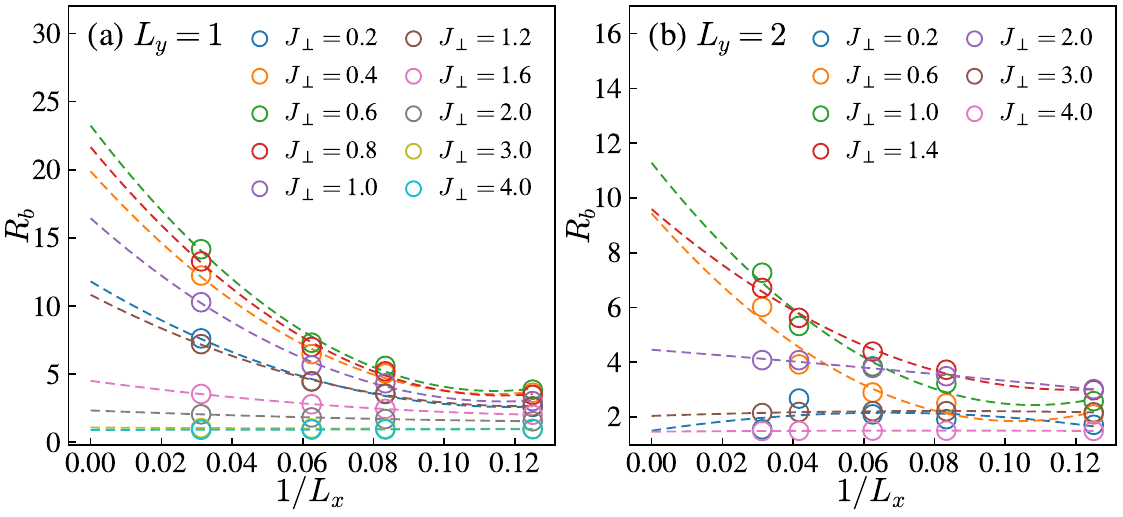}
    \caption{The extrapolation of the dimensionless ratio $R_b$ vs. system length $L_x$ by quadratic fitting based on data at $L_x=8,12,16,32$ for $L_y=1$ in panel (a) and $L_x=8,12,16,24,32$ for $L_y=2$ in panel (b).}
    \label{fig:extrapolation_rb}
\end{figure}

In the main text, the dimensionless ratio $R_b$ of binding energy is obtained by extrapolation from DMRG results via finite-size scaling with respect to $L_x$, as shown in Fig.~\ref{fig:extrapolation_rb}. The relation between the dimensionless ratio $R_b$ and system length $L_x$ is obtained by quadratic fitting $f(1/L_x)=c + b/L_x + a/L_x^2$. The curve of the dimensionless ratio vs the interlayer spin-exchange coupling is smoothed by the quadratic spline interpolation.

\begin{figure}
    \centering
    \includegraphics[width=0.7\linewidth]{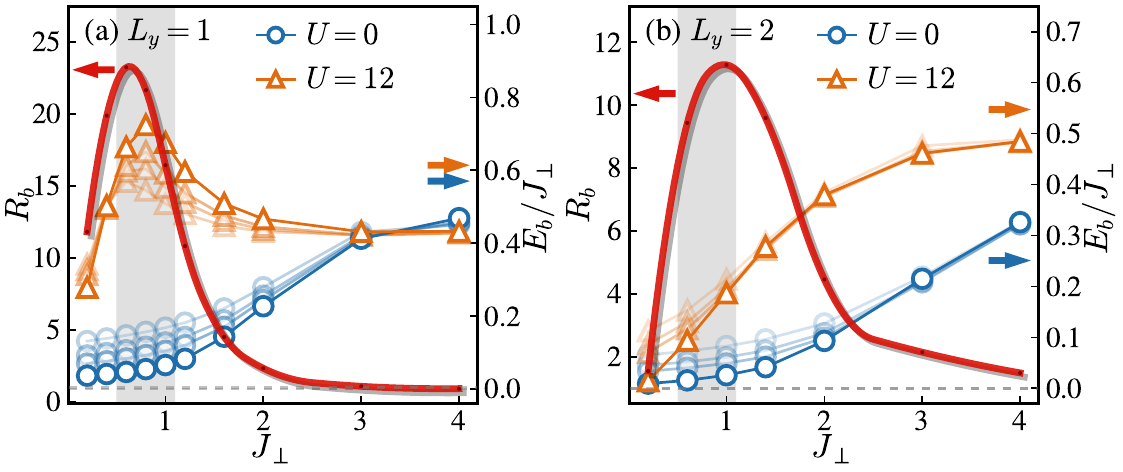}
    \caption{The binding energy $E_b$ in units of $J_\perp$ for different Hubbard repulsion $U$ and the corresponding ratio $R_b$ as a function of interlayer coupling $J_\perp$ with $L_y=1,2$ for panel (a) and (b), respectively. The system length is $L_x=8,12,16,32$ for markers of decreasing transparency. The hole doping is fixed at $\delta=1/2$. The horizontal dashed lines mark $R_b=1$ and $E_b=0$. The shaded areas indicate the experimental relevant regions of $J_\perp$.}
    \label{fig:hubbard_Eb_overJp}
\end{figure}

As $J_\perp$ directly serves as a coefficient in the Hamiltonian, the binding energy $E_b$ naturally scales linearly with $J_\perp$ in general, which makes $E_b$ look relatively small at small $J_\perp$ in FIG.~\textcolor{prlblue}{2} of the main text. Alternatively, we can depict the quantity $E_b/J_\perp$ to see the relative enhancement caused by $J_\perp$, as shown in FIG.~\ref{fig:hubbard_Eb_overJp}. One can see that the difference between $U=12$ and $U=0$ becomes larger in $E_b/J_\perp$ compared with FIG.~\textcolor{prlblue}{2} of the main text, especially in the experimental relevant region.

Here we make some clarification about the DMRG results of binding energy shown in the main text. Due to the nature of DMRG, which is believed to provide reliable results only for quasi-1D systems, it is very challenging to generalize the present numerical finding in the enhancement of the binding energy to the 2D limit. This challenge is a common and general issue for the numerical study of 2D strongly correlated systems including the single-layer Hubbard or $t$-$J$ model~\cite{Zheng2017, Qin2020}, not to mention that the bilayer structure here brings extra challenges for increasing the width of the system. However, for the current bilayer model, there is a reason to believe that the numerical results of binding energy on narrow systems can indeed be extended to wider 2D systems to a certain extent because the pairing size is quite small (around one or two lattice constants) mainly involving two holes on the two sites of a vertical bond between the two ${\rm NiO_2}$ layers~\cite{Su.Qu.2023}. Hence the pairing strength of two holes may be relatively insensitive to the width of each layer as long as the width exceeds the pairing size. 

\begin{figure}
    \centering
    \includegraphics[width=0.7\linewidth]{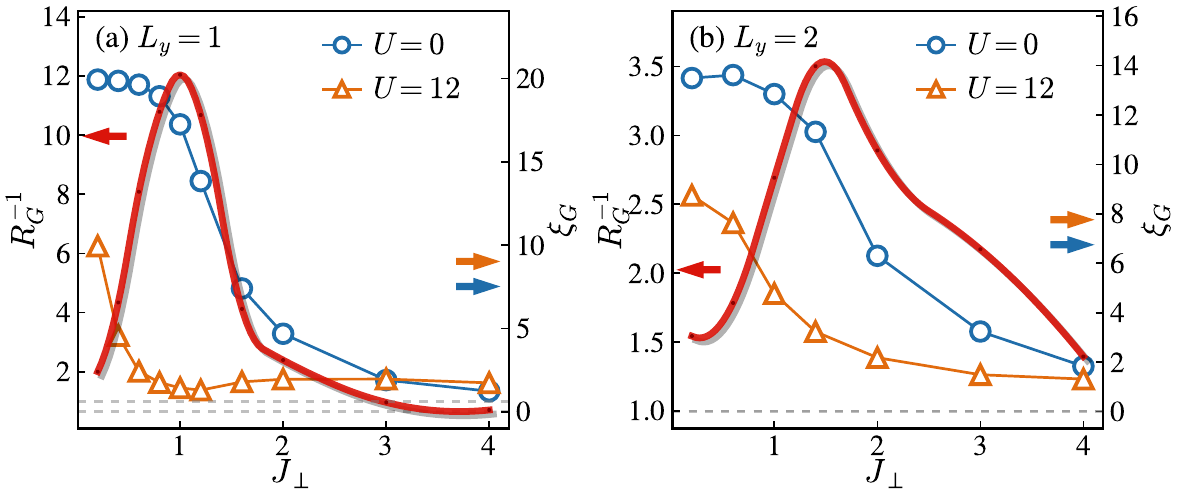}
    \caption{The correlation length $\xi_G$ extracted from the single-particle Green's function $G_{\alpha\sigma}(r)$ vs. interlayer exchange coupling $J_\perp$ for different on-site repulsion $U$ together with the inverse relative ratio $R_{G}^{-1}$. The system size is $64\times1\times 2$ for panel (a) and $32\times 2\times 2$ for panel (b), respectively.}
    \label{fig:hubbard_green_corrlen}
\end{figure}

In the main text, we show that the single-particle Green's function $G_{\alpha\sigma}(r)$ in systems with on-site repulsion $U=12$ decays exponentially for small $J_\perp$, while that with $U=0$ decays algebraically. We remark that $G_{\alpha\sigma}(r)$ is symmetric for different layer $\alpha$ and spin $\sigma$ so only the data for $G_{1\uparrow}(r)$ is depicted. Moreover, we can extract the correlation length $\xi_G$ by fitting $G_{\alpha\sigma}(r)\sim e^{-r/\xi_G}$. Although in certain cases the envelopes of $G_{\alpha\sigma}(r)$ may be closer to shapes of power-law decay, we can still extract the optimal parameter $\xi_G$ from the data of these finite-size systems to see its dependency on $J_\perp$ and $U$. As shown in Fig.~\ref{fig:hubbard_green_corrlen}, the single-particle correlation length in systems with $U=12$ is generally shorter than that with $U=0$. The inverse of the dimensionless ratio $R_G=\xi_G(U=12)/\xi_G(U=0)$ is significantly large for $J_\perp\sim t_\|$. At $J_\perp\rightarrow 0$, both cases are quasi-long-ranged but as soon as we apply a small $J_\perp$, systems with $U=12$ open a single-particle gap while systems with $U=0$ still keep the single-particle channel gapless until $J_\perp$ is large enough. This further supports the perspective that the on-site repulsion $U$ strongly frustrates the single-particle motion in the presence of a small $J_\perp$ as discussed in the main text.

\begin{figure}
    \centering
    \includegraphics[width=0.4\linewidth]{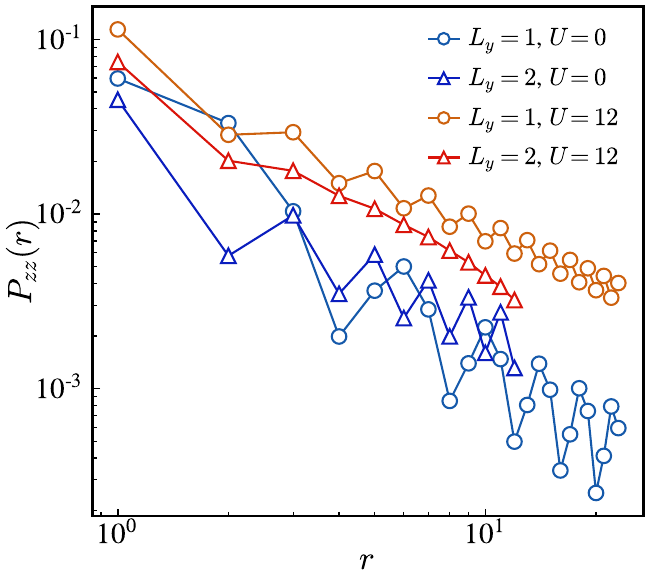}
    \caption{The pair-pair correlator $P_{zz}(r)$ for different on-site repulsion $U$ with $J_\perp/t_\|=0.6$ for $L_y=1$ and $J_\perp/t_\|=1.4$ for $L_y=2$. The system size is $64\times1\times 2$ and $32\times 2\times 2$, respectively.}
    \label{fig:hubbard_sc_zz_r}
\end{figure}

\begin{figure}
    \centering
    \includegraphics[width=0.7\linewidth]{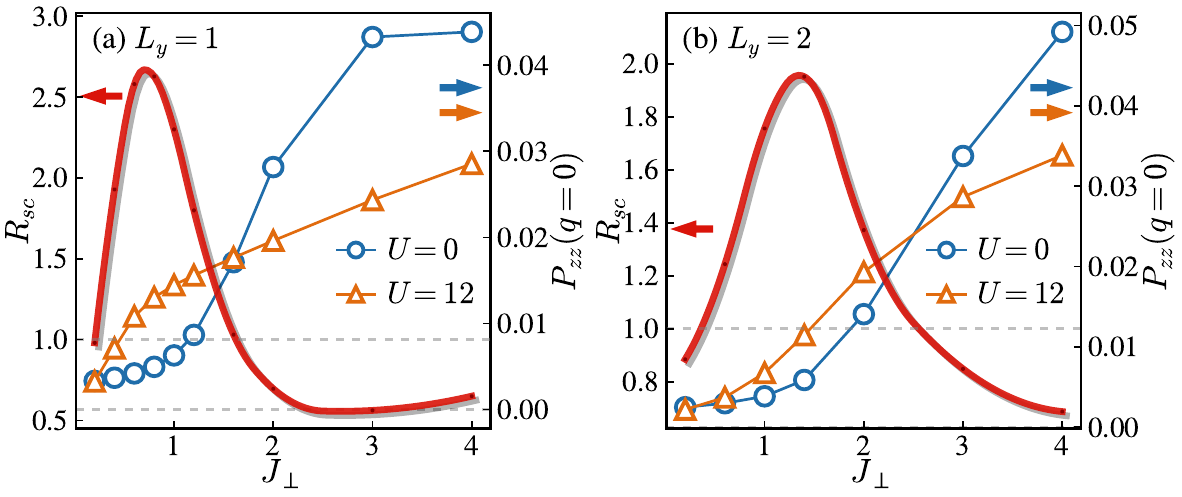}
    \caption{The overall magnitude $P_{zz}(q=0)$ of the pair-pair correlator $P_{zz}(r)$ vs. interlayer exchange coupling $J_\perp$ for different on-site repulsion $U$ together with the relative ratio $R_{sc}$. The system size is $64\times1\times 2$ for panel (a) and $32\times 2\times 2$ for panel (b), respectively.}
    \label{fig:hubbard_sc_zz_q0}
\end{figure}

In the following, we show some other correlators together with their relations with $J_\perp$ and $U$. The pair-pair correlators for singlets are defined as
\begin{equation}
    P_{\alpha\beta}(r)=\langle\Delta_{i,\alpha}^{\dagger}\Delta_{j,\beta}\rangle,
\end{equation}
respectively, where $|i-j|=r$ is the real-space distance between lattice site $i$ and $j$. Here we focus on the correlators along the $\hat{x}$ direction by taking $i=(x,y,\alpha)=(x,0,0)$ and $j=(x+r,0,0)$ and set the reference site at $x=L_x/4$. $\Delta_{i,\alpha}^\dagger= \frac{1}{\sqrt{2}}\sum_{\sigma}\sigma c_{i,\sigma}^\dagger c_{i+\alpha,-\sigma}^\dagger$ is the singlet pair creation operator defined at the bond between site $i$ and $i+\alpha$. $\alpha,\beta\in\{\hat{x},\hat{y},\hat{z}\}$ denote the bond orientation. The Fourier transform is $P_{\alpha\beta}(q)=\left|\sum_{r}P(r)e^{i q r}\right|/R$, the peak value of which can be regarded as the overall magnitude of the correlation. Fig.~\ref{fig:hubbard_sc_zz_r} shows the typical behavior of the interlayer nearest-neighbor pair-pair correlator $P_{zz}(r)$. One can see that in spite that all of them are quasi-long-ranged, $P_{zz}(r)$ in systems with $U=12$ is larger than that with $U=0$ for both $L_y=1$ and $L_y=2$. Fig.~\ref{fig:hubbard_sc_zz_q0} depicts that the overall magnitude of $P_{zz}(r)$ in systems with $U=12$ is larger than that with $U=0$ around $J_\perp\sim t_\|$, as indicated by the peak of the dimensionless ratio $R_{sc}=P_{zz}(q=0,U=12)/P_{zz}(q=0,U=0)$.

\begin{figure}
    \centering
    \includegraphics[width=0.7\linewidth]{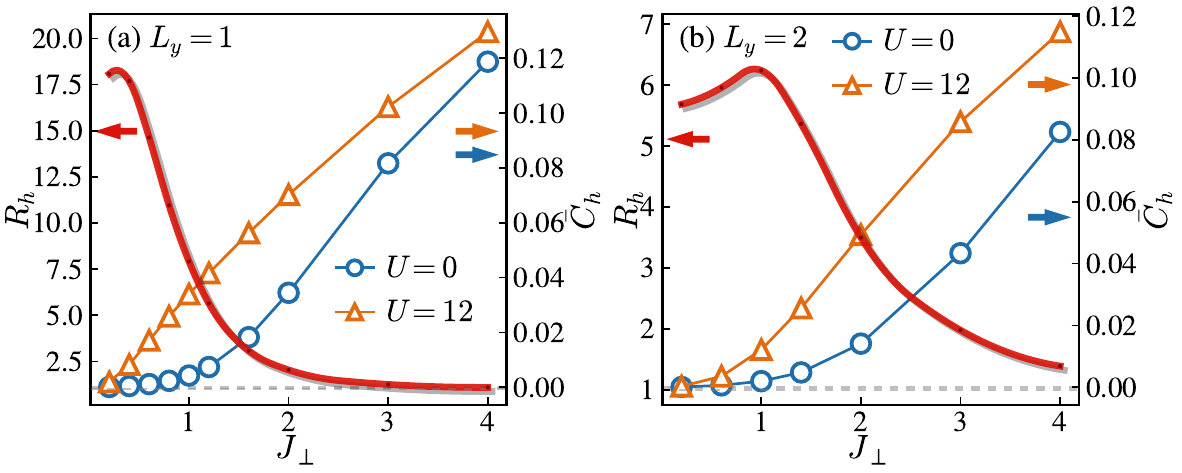}
    \caption{The averaged interlayer nearest-neighbor holon-holon correlation $\bar{C}_h$ vs. interlayer exchange coupling $J_\perp$ for different on-site repulsion $U$ in systems of (a) $L_y=1$  and (b) $L_y=2$. The system length is fixed at $L_x=32$.}
    \label{fig:hubbard_holon}
\end{figure}

We can also investigate the local pairing strength of holes through the holon-holon density correlator between interlayer nearest-neighbor sites
\begin{equation}
    C_h(i) = \langle n^h_{i1} n^h_{i2}\rangle - \langle n^h_{i1}\rangle\langle n^h_{i2}\rangle,
\end{equation}
where $n^h_{i\alpha}$ is the holon density at site $i$ and layer $\alpha$ defined by $n^h_{i\alpha}=(1-n_{i\alpha\uparrow})(1-n_{i\alpha\downarrow})$. We estimate the local pairing strength via the overall magnitude $\bar{C}_h$ by taking average over all sites, i.e., $\bar{C}_h = \sum_{i} C_h(i)/(L_xL_y)$. As shown in Fig.~\ref{fig:hubbard_holon}, $\bar{C}_h$ in systems with strong on-site repulsion $U=12$ is always larger than that with $U=0$, for both system widths $L_y=1$ and $L_y=2$. The relative ratio $R_h=\bar{C}_h(U=12)/\bar{C}_h(U=0)$ is also significantly large at $J_\perp\lesssim t_\|$. This implies that given a holon at site $i$ and layer $1$, it is more likely to find another holon at site $i$ and layer $2$ in systems with $U=12$ than those with $U=0$. This data provides additional evidence for the enhancement of interlayer pairing strength from the on-site repulsion, compensating for the deficiency that the binding energy shown in the main text mixes the contributions from interlayer and intralayer pairing.

\end{document}